# Giant inverse magnetocaloric effect near room temperature in Co substituted NiMnSb Heusler alloys

**Ajaya K. Nayak**[1]**, K. G. Suresh**[1]**and A. K. Nigam**[2]

[1]Magnetic Materials Laboratory, Department of Physics, Indian Institute of Technology Bombay, Mumbai-400076, India.

[2]Tata Institute of Fundamental Research, Homi Bhabha Road, Mumbai-400005, India

E-mail: suresh@phy.iitb.ac.in

The effect of Co on the structural, magnetic and magnetocaloric effect (MCE) of $Ni_{50-x}Co_xMn_{38}Sb_{12}$ (x=0,2,3,4,5) Heusler alloys was studied. Using x-ray diffraction, we show the evolution of the martensitic phase from the austenite phase. The martensitic transition temperature is found to decrease monotonically with Co concentration. Remarkable enhancement of MCE is observed near room temperature upon Co substitution. The maximum magnetic entropy change of 34 $Jkg^{-1}K^{-1}$ was achieved in x=5 at 262 K in a field of 50 kOe and a value of 29 $Jkg^{-1}K^{-1}$ found near room temperature. The significant increase in the magnetization associated with the reverse martensitic transition is responsible for the giant MCE in these compounds.

Key words: Magnetocaloric effect, Heusler alloys, martensitic transition

# 1. Introduction

The search for an alternative to the conventional gas-based refrigeration technique has led to the discovery of magnetic refrigeration, which has become very promising recently. The growth of this technique depends on the availability of potential solid state magnetic materials, which act as refrigerants. The underlying property of magnetic refrigerants is the magnetocaloric effect (MCE), which is measured in terms of isothermal magnetic entropy change ($\Delta S_M$) and/or adiabatic temperature change ($\Delta T_{ad}$). In recent years, there have been a large number of studies in various magnetic materials for finding out their suitability as potential refrigerants. Large MCE is generally observed in those materials which undergo magneto-structural transition and/or first order magnetic transitions [1,2]. The first and foremost system to show giant MCE is $Gd_5(Si_{1-x}Ge_x)_4$. Following this discovery, a few other systems such as Mn-Fe-As, $La(Fe,Si)_{13}$ etc. were found to exhibit giant MCE [1-5].

Recently there has been a great interest in Ni-Mn based Heusler alloys because of the observation of first order martensitic transformation around room temperature. These alloys received a lot of attention for showing both magnetic shape memory effect and large magnetocaloric effect [6-8]. Among the Heusler alloys, the highest MCE has been reported in $Ni_{55}Mn_{20}Ga_{25}$ single crystals ($\Delta S_M$= -86Jkg$^{-1}$K$^{-1}$ in a field of 50kOe), which is due to the coincidence of first order structural transition and the magnetic transition [7]. Other materials showing promising MCE from the Ni-Mn family are Ni-Mn-Sn [9], Ni-Mn-In [10] and Ni-Mn-Sb [11,12]. All these alloys generally show inverse (negative) MCE, *i.e.*, the entropy change is positive on applying the magnetic field. It is reported that Ni-Mn-Sb based Heusler alloys undergo first order magneto-structural transition from martensite phase (tetragonal or orthorhombic) to austenite phase (cubic), resulting in a large MCE of 19 Jkg$^{-1}$K$^{-1}$ for a field of 50 kOe [11]. Recently, several studies have also reported the magneto-structural properties of Ni-Mn-Sb system [13-15]. With the aim of developing novel and better materials belonging to this series and to probe the effect of structural variation on the MCE in them, we have partially substituted Co for Ni in the non-stoichiometric NiMnSb alloy, resulting in the series Ni$_{50-x}$Co$_x$Mn$_{38}$Sb$_{12}$. In this paper, we report the variations in the structural, magnetic and magnetocaloric properties of the alloys with $x$=0, 2, 3, 4 and 5.

## 2. Experimental Details

Polycrystalline ingots of Ni$_{50-x}$Co$_x$Mn$_{38}$Sb$_{12}$ were prepared by arc-melting the stoichiometric amounts of Ni, Co, Mn, Sb of at least 99.99% purity in high pure argon atmosphere. The ingots were re-melted four times and were subsequently annealed in evacuated quartz tubes at 850˚C for 24 hours. The structural characterization of the materials was done by powder x-ray diffractograms (XRD) using Cu-Kα radiation. The magnetization measurements were carried out using a vibrating sample magnetometer attached to a Physical Property Measurement System (Quantum Design, PPMS-6500).

## 3. Results and discussion

XRD patterns show that the alloys with $x$=0, 2 and 3 show the martensitic (orthorhombic) phase while the ones with $x$=4 and 5 show the austenite (cubic) phase, at room temperature. The Rietveld refinement has shown that all the alloys are single phase at room temperature and that the lattice parameter variation with Co concentration is quite negligible. In order to find out the evolution of the martensitic phase from the austenite phase with reduction in temperature, we have focused on the alloy with $x=4$, which is austenite at room temperature. In this alloy, we have carried out the XRD study at various temperatures $\leq 320$ K. Figure. 1 shows how the original L2$_1$ (220)$_{cubic}$ peak of the austenite phase splits into various 10M modified orthorhombic peaks corresponding to the martensitic phase with decreasing temperature.

The temperature dependence of magnetization ($M$) in Ni$_{50-x}$Co$_x$Mn$_{38}$Sb$_{12}$ alloys with $x=3$ and 5 is shown in Figure. 2. The measurements have been performed in an applied field of 1 kOe, in three different modes. In the first case, the sample was initially cooled to 5 K without applying any magnetic field and then data was taken as the temperature was increased from 5 K by applying 1 kOe (ZFC). In the FC mode, the data was collected during the cooling, while in the FH mode, the data was collected while heating, after field cooling. In the latter two modes also, the applied field was 1 kOe. The ferromagnetic-paramagnetic transition of the austenite phase ($T_C^A$) occurs nearly at the same temperature (~330 K) for all the compounds. As the temperature is reduced below $T_C^A$, the magnetization sharply decreases, after reaching a maximum value. The temperature corresponding to this maximum point is called the martensitic start ($M_S$) and the minimum point called the martensitic finish temperature ($M_F$), as indicated in Figure. 2. Just below $M_S$, the austenite phase loses stability and the magnetization decreases in this mixed austenite-martensite region. At temperatures slightly below $M_F$, an additional transition is observed, which corresponds to the Curie temperature ($T_C^M$) of the martensitic phase. Across the series, $M_S$ is found to decrease from 330 K to 270 K and $T_C^M$ is found to decrease from 275 K to 235 K, as $x$ is increased from 0 to 5. By correlating the *M-T* and the XRD data of all the five samples, it can be seen that the martensitic phase gets destabilized with Co concentration.

Another feature worth noting in Figure 2 is the splitting between the ZFC and FC data at low temperatures (in the martensitic phase), which indicates the presence of magnetic frustration. It has been reported that in off-stoichiometric NiMnSn alloys, a part of the Mn ions occupies the Sn site. The magnetic coupling between the Mn ions occupying the Mn and Sn sites is antiferromagnetic (AFM), while that between the Mn ions occupying the regular Mn site is ferromagnetic (FM) [16]. A similar scenario may exist in the case of NiMnSb as well. There may also exist a FM component at low temperatures, as in the case of NiMnIn [17]. Domain wall pinning of the FM component due to the larger anisotropy of the non-cubic (martensitic) phase may also contribute to the thermomagnetic irreversibility seen between the FC and ZFC data. On the other hand, in all the compounds studied, the difference between FC and FH curves reflects the thermal hysteresis associated with the first order structural transition. In all the alloys, the ZFC and FH data are found to coincide, except at low temperatures. M-T data seen in the three modes in the present case is similar to the one reported by Aksoy et al. in Ni-Mn-In(Ga) Heusler alloys [18]. It is also found that the $M_S$ shifts towards lower temperatures with increase in field, which implies that the field also tries to suppress the martensitic transition. In the case of x=5, $M_S$ is 270 K in a field of 1 kOe and decreases to 252 K at 80 kOe, as is evident from Figure 3.

Figure 4 shows the magnetization isotherms of $Ni_{45}Co_5Mn_{38}Sb_{12}$. At 252 K and 276 K it shows the simple ferromagnetic behavior. The metamagnetic-like character in the *M-H* isotherms at 261 K (just below $M_S$) is associated with a field-induced reverse martensitic transformation. At this temperature, with increasing field, the magnetization initially increases rather slowly, but above a certain field, it begins to increase faster. This is because of the field induced transition from low magnetization martensitic state to a higher magnetization austenite state.

.

The magnetocaloric effect ($\Delta S_M$) was calculated from the isothermal magnetization curves, close to the martensitic/magnetic transition temperatures. The isotherms have been obtained in an interval of 3 K, in the increasing temperature mode and the isothermal magnetic entropy was calculated using the Maxwell's relation [2,9] given by

$\left(\dfrac{\partial S(T,H)}{\partial H}\right)_T = \left(\dfrac{\partial M(T,H)}{\partial T}\right)_M$, which gives

$$\Delta S_M(T,H) = \int_0^H \left(\dfrac{\partial M(T,H)}{\partial T}\right)_H dH \tag{1}$$

Since we used a constant temperature interval, equation (1) can be written as

$$\Delta S_M \approx \dfrac{1}{\Delta T}\left[\int_0^{H'}(M(T+\Delta T,H) - M(T,H))dH\right] \tag{2}$$

Figure 5 shows the $\Delta S_M$ as a function of temperature for various applied fields ($\Delta H$) for all the compounds. The entropy change shows a spike-like behavior as a function of temperature. It can be seen that the entropy change is positive and that the $(\Delta S_M)_{max}$ increases with field in all the compounds. Among the compounds studied in this work, the highest $(\Delta S_M)_{max}$ of 34 Jkg$^{-1}$K$^{-1}$ has been observed for x=5. The value of 29 Jkg$^{-1}$K$^{-1}$ in x=4 is also very promising due to the fact that it occurs near room temperature. It may be noted that $(\Delta S_M)_{max}$ in Ni$_{50}$Mn$_{38}$Sb$_{12}$ is only about 7 Jkg$^{-1}$K$^{-1}$, which increases monotonically with Co concentration, as shown in the inset of Figure 5. These values are also much larger than the reported value of 9.1 Jkg$^{-1}$K$^{-1}$ in Ni$_{50}$Mn$_{37}$Sb$_{13}$ [12]. We have also observed that the entropy change associated with the magnetic transition at $T_C^A$ is negative and the magnitude is comparatively much smaller (~3 J/kgK).

The possibility of tuning the martensitic transition temperature and the observation of large MCE at temperatures close to room temperature make this system quite interesting from the points of view of fundamental aspects as well as applications. Co is found to stabilize the austenite phase, resulting in the shift of the martensitic transition to lower temperatures. This is due to the decrease in the (e/a) ratio upon Co substitution. The enhancement of MCE with Co substitution in this system is quite remarkable in the light of the general observation that the MCE decreases with Co substitution (for Ni) in Ni-Mn based Heusler alloys. For example, Co substitution is found to reduce the entropy change from 20 J/kg K in Ni$_{50}$Mn$_{37}$Sn$_{13}$ to 9 J/kg K in Ni$_{47}$Co$_3$Mn$_{37}$Sn$_{13}$ [8]. A similar decrease is reported in Ni-Co-Mn-Ga as well [19].

Since Co has a larger magnetic moment (~1 $\mu_B$) as compared to the Ni (0.3 $\mu_B$), substitution of Co for Ni would increase the ferromagnetic coupling in this system. This is reflected in the observation that the magnetization (at 80 kOe) just above $M_S$ increases from 22 emu/g for x=0 to 65 emu/g for x=5. On the other hand, the magnetization is found to decrease monotonically with Co, in the martensitic phase. Therefore, the large positive magnetic entropy change is related to the significant increase in ($\partial M / \partial T$) value with Co concentration. Enhanced magneto-structural coupling brought about by Co is responsible for this variation of the magnetization in the austenite and martensitic phases. The fact that the hysteresis between the FC and FH curves increases with Co also justifies this proposition. By comparing the MCE variation in Co substituted NiMnSn and the present series, we feel that the magneto-elastic coupling induced by Co in the latter case is quite large compared to that in the former. In this context, it is of importance to note that Yu et al. have recently observed a large magnetoresistance of about 70% in $Ni_{41}Co_9Mn_{39}Sb_{11}$ alloy in a field of 100 kOe[20].

## 4. Conclusions

In conclusion, we have studied the structural variation of NiCoMnSb alloys by combining the XRD and magnetization studies. Increase in Co concentration leads to a gradual decrease in the martensitic transition temperature and a large increase in the MCE. The increase in MCE is attributed to the large magneto-structural coupling, which causes a large $\partial M / \partial T$ upon Co substitution.

## Acknowledgement

KGS and AKN thank the B.R.N.S., Govt. of India for financially supporting this work.

**Figure captions**

Figure.1. Temperature variation of XRD patterns for $Ni_{46}Co_4Mn_{38}Sb_{12}$, showing the change in phase from austenite to martensite.

Figure.2. Zero field cooled (ZFC), field cooled (FC) and field heated (FH) magnetization data as a function of temperature for $Ni_{50-x}Co_xMn_{38}Sb_{12}$ alloys with $x=3$ and 5 in a field of 1 kOe. Inset shows the change in magnetization with temperature near the martensitic transition temperature.

Figure.3 Temperature dependence of magnetization for $Ni_{45}Co_5Mn_{38}Sb_{12}$ in different fields.

Figure.4. Magnetization isotherms of $Ni_{45}Co_5Mn_{38}Sb_{12}$ in various temperature ranges.

FIigure.5. $\Delta S_M$ as a function of temperature for $Ni_{50-x}Co_xMn_{38}Sb_{12}$ alloys with $x=0, 2, 3, 4$ and 5 in fields of 20 and 50kOe. The inset shows the variation of $(\Delta S_M)_{max}$ with $x$.

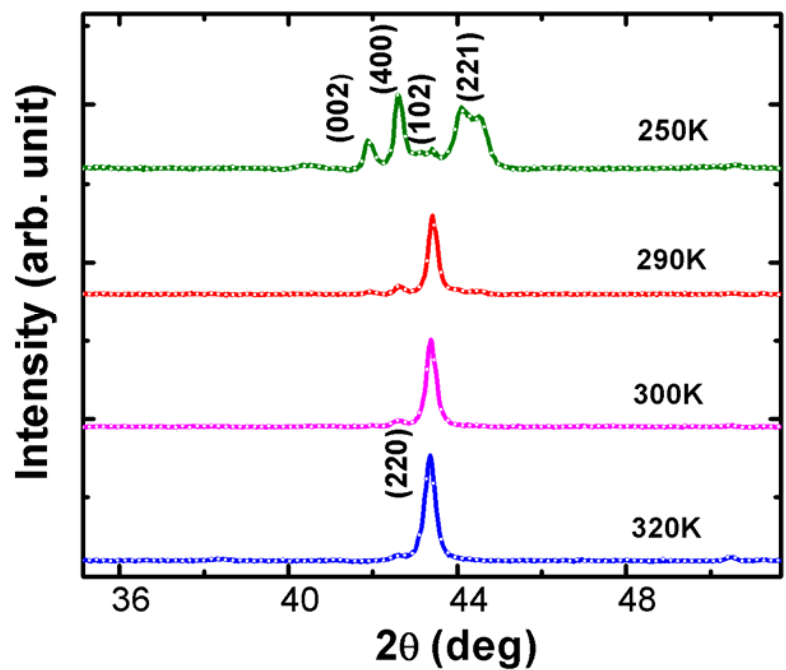

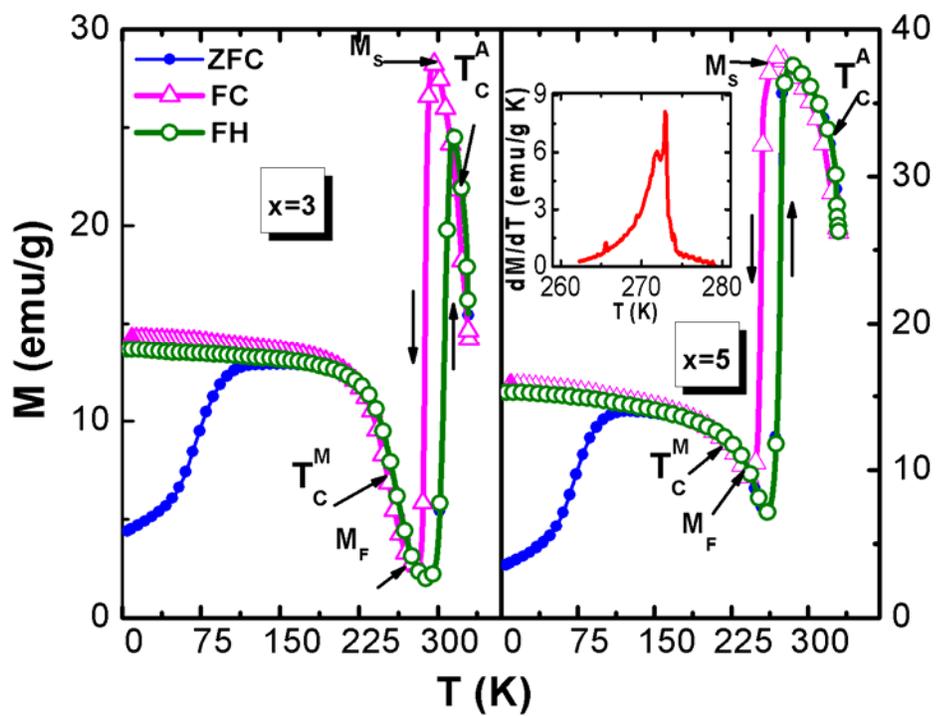

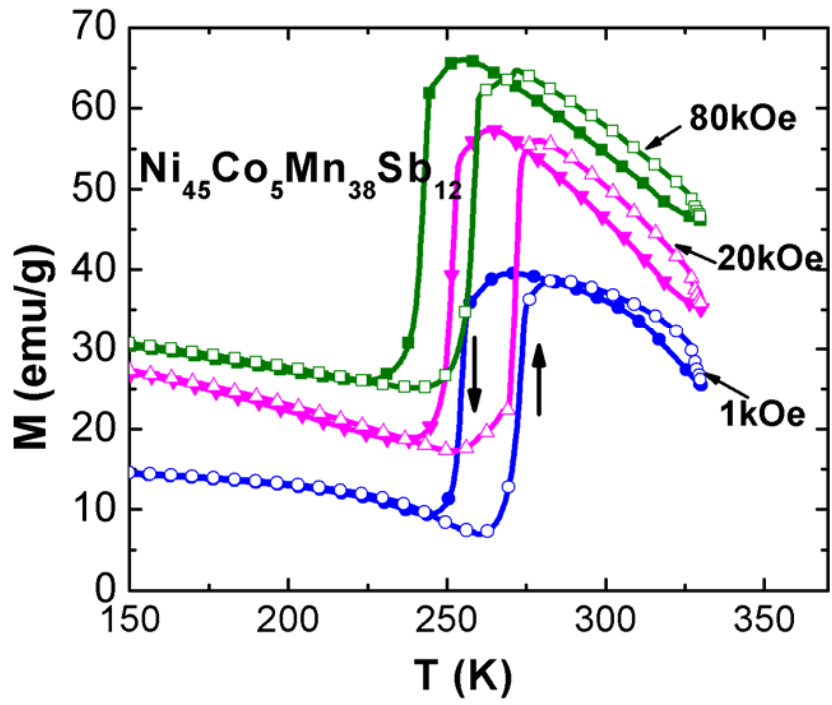

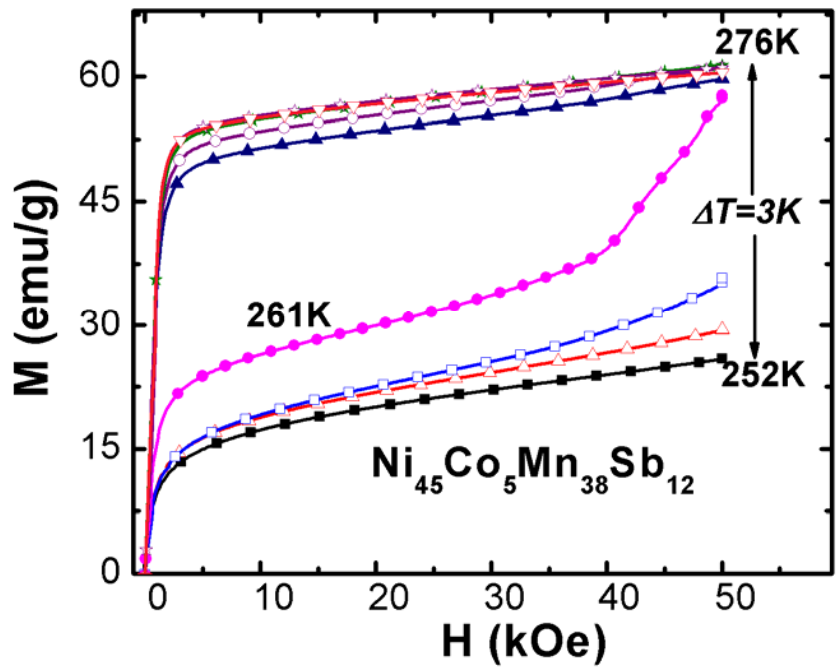

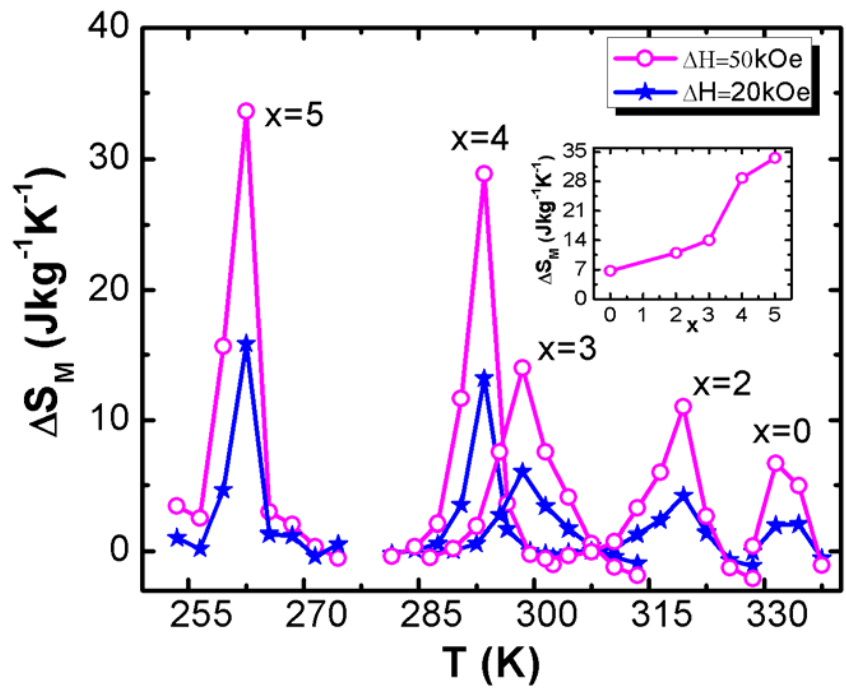